\documentclass[apj,numberedappendix,tighten]{emulateapj}

\pdfoutput=1

\usepackage{amsmath}
\usepackage{rotating}

\PassOptionsToPackage{breaklinks=true}{hyperref}


\def\***#1{\textbf{\boldmath\textsf{***#1***}}}

\journalinfo{Accepted for publication in The Astrophysical Journal}
\shorttitle{NGC 5044 Thermally Unstable X-Ray Filaments}
\shortauthors{DAVID ET AL.}

\begin{document}

\title{The Presence of Thermally Unstable X-ray Filaments and the Production of Cold Gas in the NGC 5044 Group}

\author{Laurence P. David$^1$, Jan Vrtilek$^1$, Ewan O'Sullivan$^1$, Christine Jones$^1$, William Forman$^1$ \& Ming Sun$^2$}

\affil{$^1$Harvard-Smithsonian Center for Astrophysics, 60 Garden St., Cambridge, MA 02138, USA}
\affil{$^2$Department of Physics, University of Alabama in Huntsville, Huntsville, AL 35899, USA}

\begin{abstract}
We present the results of a deep {\it Chandra} observation of the X-ray
bright, moderate cooling flow group NGC 5044 along with the
observed correlations between the ionized, atomic, and molecular gas
in this system.  The {\it Chandra} observation shows that the central
AGN has undergone two outbursts in the past $10^8$~years, based on the presence of
two pairs of nearly bipolar X-ray cavities.  The molecular gas
and dust within the central 2~kpc is aligned with the orientation of the
inner pair of bipolar X-ray cavities, suggesting that the most recent
AGN outburst had a dynamical impact on the molecular gas.
NGC 5044 also hosts many X-ray filaments within the central 8~kpc,
but there are no obvious connections between the X-ray
and H$\alpha$ filaments and the more extended X-ray cavities that were inflated
during the prior AGN outburst.  Using the linewidth of the blended Fe-L line complex
as a diagnostic for multiphase gas, we find that the majority of
the multiphase, thermally unstable gas in NGC 5044 is confined within the
X-ray filaments.  While the cooling time and entropy of the gas within the
X-ray filaments are very similar, not all filaments show evidence of gas cooling
or an association with H$\alpha$ emission.
We suggest that the various observed properties of the X-ray filaments are
suggestive of an evolutionary sequence where thermally unstable gas begins to cool,
becomes multiphased, develops H$\alpha$ emitting plasma, and finally
produces cold gas.

\end{abstract}


\section{Introduction}

Groups and clusters of galaxies have been traditionally divided
into two main categories: cool-core systems, which have smooth X-ray
morphologies and cooler gas in their centers; and non-cool-core
systems, which have complex X-ray morphologies and no cool gas.
Early X-ray observations showed that the radiative cooling time of gas
in the center of cool-core clusters can be significantly shorter than
a Hubble time, and that in the absence of reheating, the gas
should cool at rates of up to hundreds of M$_{\odot}$~yr$^{-1}$ (Fabian et al. 1984).
The primary issue with the early cooling flow scenario was the
lack of significant amounts of cold gas or star formation.

Our understanding of cooling flows changed dramatically
with the launch of the {\it Chandra} and {\it XMM-Newton} X-ray observatories when
a strong feedback mechanism was discovered between the central AGN
and hot gas that prevents the bulk of the hot gas from cooling
(e.g., McNamara et al. 2000; David et al. 2001; Fabian et al. 2003a;
Blanton et al. 2003; Peterson \& Fabian 2006; Forman et al. 2007;
McNamara \& Nulsen 2011 and references therein).
The AGN-cooling flow feedback mechanism has far-reaching
consequences concerning galaxy formation and can explain the observed
correlation between bulge mass and black hole mass
(Magorrian et al. 1998; Gebhardt et al. 2000) and the cutoff in
the number of massive galaxies.

Over the same time period, many studies
have shown that the central dominant galaxies (CDGs) in cooling flows
are far from being red and dead. Single-dish CO
surveys over the past decade have shown that a substantial fraction of
CDGs in groups and clusters harbor large amounts of molecular gas
(Edge 2001; Salome \& Combes 2003; Combes et al. 2007; Young et al. 2011).
A great deal of more detailed information about the molecular gas content
in cooling flows has been acquired recently with ALMA
(McNamara et al. 2014; Russell et al. 2014; David et al. 2014;
Russell et al. 2016,2017).  In addition, star formation has now been detected in
many CDGs in cooling flows with rates roughly consistent with
the AGN-suppressed mass-cooling rates (Rafferty et al. 2006; Quillen et al. 2008).
The fact that star formation and emission from cooler gas
(e.g., H$\alpha$, [CII], and CO) are only detected in systems with
short cooling times (Rafferty et al. 2008) or low gas entropies
(Cavagnolo et al. 2008,2009) confirms that the
hot gas is the primary reservoir fueling star formation.  Cooling flows
have always been a study of star formation in a hostile environment, but
only recently has it become possible through the combination of multifrequency
data to trace all the stages of star formation.

A wealth of multifrequency data is available for the cooling group NGC 5044,
which makes it an ideal object for studying correlations between gas over
a broad range of temperatures.  Previous X-ray studies of NGC 5044 using
{\it XMM-Newton} and earlier {\it Chandra}
data show that the hot gas within the central region of NGC 5044 has been
perturbed by AGN outbursts and the motion of the central galaxy within the
group potential, based on the presence of X-ray filaments, cavities, and cold fronts
(Buote et al. 2003; David et al. 2009,2011,
Gastaldello et al. 2009,2013; O'Sullivan et al. 2014).
David et al. (2009) showed that mechanical heating by AGN-inflated cavities
is sufficient to offset radiative cooling of the gas within the central
10~kpc. While AGN-feedback is sufficient to prevent the bulk of gas from cooling
in the center of NGC 5044, some gas must be cooling out of the hot phase to supply
the cooler gas detected in multifrequency observations, including bright
H$\alpha$ filaments, ro-vibrational H$_2$ line emission (Panuzzo et al. 2011),
[CII] line emission (Werner et al. 2014), and CO emission (David et al. 2014).
Ongoing star formation also has been observed in NGC 5044 at a rate of
about 1\% of the classical cooling rate.  The primary objective of this paper is
to identify the location of the small fraction of hot gas that is actually
cooling out of the hot phase and to search for correlations between the ionized,
atomic, and molecular gas.

\begin{deluxetable}{rccc}
\tablewidth{3.0in}
 \tablecaption{NGC 5044 Observing Log}
 \tablehead{
 \colhead{Date} &
 \colhead{ObsID} &
 \colhead{Total Exposure} &
 \colhead{Cleaned Exposure} 
 \\
 \colhead{} &
 \colhead{} &
 \colhead{(ks)} &
 \colhead{(ks)} 
 }
\startdata
2000 Mar 19 &  798  & 20.5 & 18.9  \\
2008 Mar  7 &  9399 & 82.7 & 80.4  \\ 
2015 May  7 & 17653 & 35.5 & 32.6  \\ 
2015 May 10 & 17654 & 25.0 & 23.5  \\ 
2015 May 11 & 17196 & 88.9 & 83.8  \\ 
2015 Jun  6 & 17195 & 78.0 & 74.5  \\ 
2015 Aug 23 & 17666 & 88.5 & 81.9  \\ 
\enddata
\tablecomments{Observation date, Chandra observation ID number, total exposure time,
and cleaned exposure time for all ACIS observations of NGC 5044.}
\end{deluxetable}

This paper is organized as follows.  Section 2 contains a description
of the {\it Chandra} data reduction. An overview of all multifrequency
data is presented in $\S 3$ and the global X-ray properties of NGC 5044
are given in $\S 4$. In $\S 5$ we identify the
location of the multiphase, thermally unstable gas and show that
the observed properties of the X-ray filaments are consistent with
the cooling of thermally unstable gas. The origin of the
molecular gas kinematics is examined in $\S 6$ and correlations
between the hot gas, cold gas, and star formation are discussed
in $\S 7$.  Finally, our main results are summarized in $\S 8$.

\section{{\it Chandra} Data Reduction}

An observing log of all {\it Chandra} observations of the
NGC 5044 group is shown in Table 1.
All observations were centered on the backside-illuminated ACIS-S3 chip
and reprocessed with CIAO 4.8 and CALDB 4.7.2.
Except for the 2000 observation, all observations were carried out in very
faint (VF) telemetry format, and VF background filtering was applied to the
appropriate data sets.
Since the X-ray emission from NGC 5044 covers the entire S3 chip,
data acquired on the backside-illuminated S1 chip, which was
turned on during all the observations, was used to screen for background flares.
Light curves were generated in the 3.0-8.0~keV and 9.0-12.0~keV energy
bands for the diffuse emission on S1, and time intervals with
count rates exceeding 20\% of the mean count rate were excluded from further analysis.
Background data sets were generated for all active chips during the ACIS
observations by extracting the appropriate blank sky background files from
the {\it Chandra} CALDB based on the date of the observations.
VF filtering was then applied to the background data sets, and
the live time in each background file was adjusted to yield the
same 9.0-12.0~keV count rate as in the corresponding NGC 5044
observation.

The CIAO tool {\it wavdetect} was run on a mosaic of all seven ObsIDs
to identify point sources, which were then excised from the analysis of
the diffuse emission in the NGC 5044 group.  Point sources were detected
in a broad energy band from 0.5-7.0~keV and a hard energy band from
2.0-7.0~keV.  The broad energy band is sufficient to detect point
sources outside of the core of NGC 5044, but a harder energy band
is required to detect the emission from the host LMXBs in NGC 5044 because
of the the high surface brightness of the hot gas (see David et al. 2009).

As shown below, there is significant structure in the hot gas within the central
few arcseconds of the NGC 5044 group.  After reprocessing all of the ACIS data,
the rms scatter in the X-ray centroid of the AGN among the seven ObsIDs
was $0^{\prime\prime}.53$ (within the quoted uncertainties in the absolute
astrometry of {\it Chandra} data\footnote{cxc.harvard.edu/proposer/POG/}).
To obtain the best possible imaging within this region, we used the X-ray centroid
of the AGN as derived
from ObsID 17196 (the deepest exposure) as the fiducial value,
and re-registered the remaining ObsIDs to produce consistent X-ray centroids for
the AGN using the CIAO science thread,
{\it Correcting Absolute Astrometry}.\footnote{cxc.harvard.edu/ciao/threads/reproject\_aspect/}
We adopt a luminosity distance of 31.2~Mpc to NGC 5044 (Tonry et al. 2001), which gives a
physical scale of $1^{\prime\prime}=150$~pc.
All uncertainties are quoted at the $1~\sigma$ level unless otherwise noted.

\begin{figure}[t]
\center{\includegraphics[width=1.00\linewidth,bb=44 144 492 588,clip]{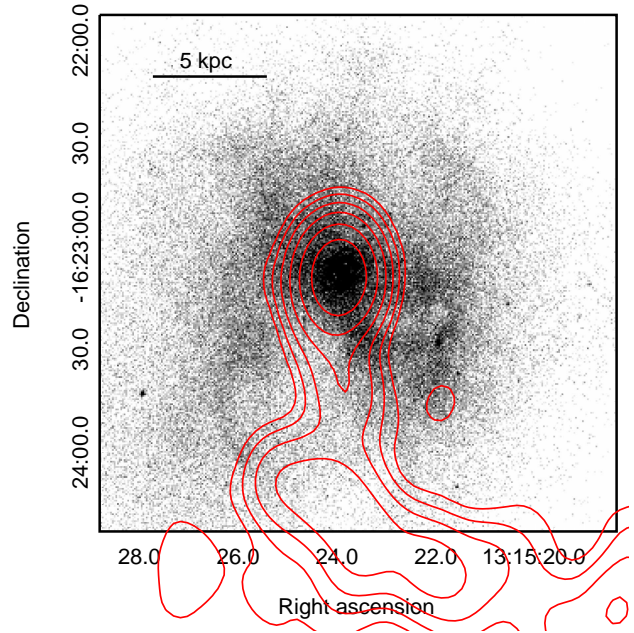}}
\caption{Raw 0.5-2.0~keV image of the combined 395~ks of {\it Chandra} ACIS data within
the central 22.5 by 22.5~kpc region along with 235~MHz contours from our GMRT observation
(see Giacintucci et al. 2011 for further information about the GMRT data).}
\end{figure}

\section{MultiFrequency Overview}

A mosaic of the combined 395~ks of ACIS data is shown in Figure 1, which highlights
the X-ray structures on scales of a few kpc.  Also shown in Figure 1 are the 235~MHz radio
contours from our GMRT observation (Giacintucci et al. 2011).
The X-ray cavities and filaments on these scales are the most prominent in the
unsharp masked image shown in Figure 2.  The largest X-ray cavity extends from
slightly SE of the AGN toward the south and is filled with 235~MHz radio
emission. There is also an X-ray cavity toward the NW in Figure 2 that
was probably inflated by the same AGN outburst, but no 235~MHz emission was
detected within this region above the sensitivity limit in our GMRT observation.

Many other X-ray cavities are also apparent in the unsharp masked image.
To determine if the lower surface brightness regions seen in Figure 2 are
true AGN-inflated cavities, we extracted tangential surface brightness profiles within
the annulus shown in Figure 2 in both a soft and hard energy band (see Figure 3).
The NW and SE cavities are both easily identifiable in this figure.  The surface
brightness depressions within the NW and SE cavities are also more significant
in the soft band, indicating that these cavities were produced by the evacuation
of the cooler central gas, and not hotter gas at larger radii seen in projection.
In contrast, all other apparent cavities seen in Figure 2 have surface brightnesses
in both energy bands consistent with the azimuthally averaged surface brightness
within the annulus.  This result shows that there are only two AGN-inflated
cavities on intermediate scales in NGC 5044, probably produced by a single AGN outburst,
and that the other apparent X-ray cavities seen in Figure 2 are simply average
surface brightness regions surrounded by X-ray bright filaments.

\begin{figure}[t]
\center{\includegraphics[width=1.00\linewidth,bb=53 163 561 632,clip]{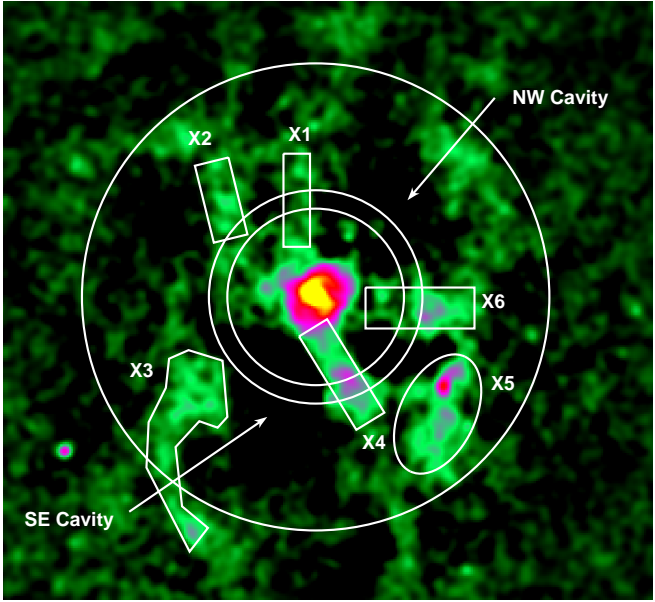}}
\caption{Unsharp masked 0.5-2.0~keV image (the difference between gaussian smoothed
images at $2^{\prime\prime}$ and $20^{\prime\prime}$) of the combined
395~ks of ACIS data. The brightest filaments are identified in the 
figure. Also shown are the locations of the southeastern and northwestern 
cavities and the annulus used to extract the tangential surface brightness profile 
shown in Figure 3. The outer circle has a radius of 8~kpc.}
\end{figure}

The X-ray bright filaments (X1, X2, X4, and X6) that intersect the annulus shown in
Figure 2 are also easily identifiable as enhancements in the surface brightness profiles
shown in Figure 3. However, the X1 and X4 filaments are much more pronounced in the soft
band than the hard band, while the X2 and X6 filaments show the same
enhancement in both energy bands. This suggests that either the X1 and X4 filaments
are simply cooler than the X2 and X6 filaments or they contain multiphase gas.
A full spectroscopic analysis of the X-ray filaments is presented in $\S 5.2$.

\begin{figure}[t]
\center{\includegraphics[width=1.00\linewidth,bb=20 145 576 698,clip]{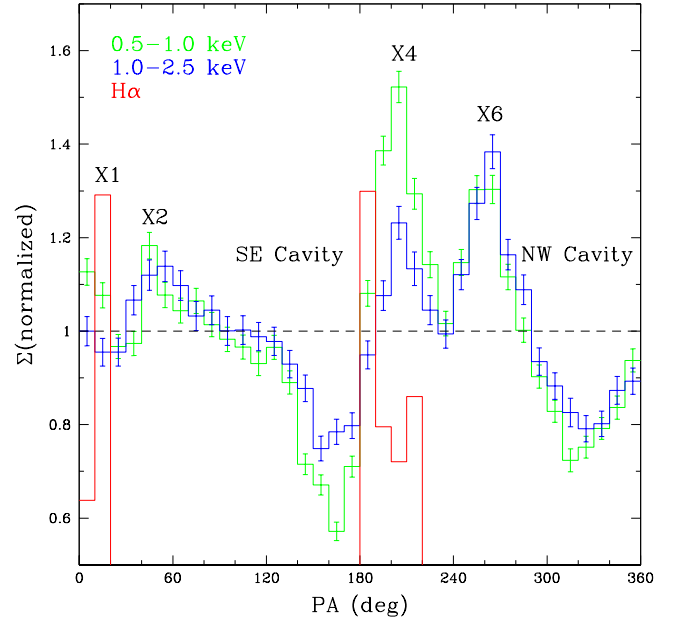}}
\caption{Tangential surface brightness profile within the annulus shown
in Figure 2.  The position angle (PA) is measured counterclockwise from 
due north.  Profiles are shown in a soft (0.5-1.0~keV) band (green) and a hard
(1.0-2.5~keV) band (blue).  The two X-ray surface brightness profiles are normalized to the 
mean value within the annulus. The tangential H$\alpha$ surface brightness profile (red)
is also shown with an arbitrary normalization. The locations of the 
X-ray cavities and filaments identified in Figure 2 are also shown.}
\end{figure}

NGC 5044 hosts a very luminous system of H$\alpha$ filaments with
$\rm{L}_{H \alpha} = 4.6 \times 10^{40}$~erg~s$^{-1}$ (Werner et al. 2014).
Figure 4 shows a comparison between the X-ray and H$\alpha$ filaments.
Note that all of the X-ray and H$\alpha$ filaments lie in the region between
the two largest X-ray cavities.  In addition, only the X1 and X4 filaments,
which have the softest X-ray spectra, are coincident with extended H$\alpha$
filaments, while the remaining X-ray filaments only contain diffuse or very
faint filamentary H$\alpha$ emission.  We also show a comparison between
X-ray and H$\alpha$ emission in Figure 3 where we include the tangential H$\alpha$
surface brightness profile within the same annulus as used for the X-ray data.
Only the X1 and X4 filaments are associated with H$\alpha$ emission in Figure 3;
however, the brightest portion of the H3 filament is slightly offset
from the brightest portion of the X4 filament.  This is further illustrated in
Figure 5 which shows that the H3 filament lies along the interface between the
X4 filament and the southeastern cavity.

By combining our deep {\it Chandra} observation with archival data,
we are able to investigate the central region of NGC 5044 on arcsecond scales,
which was not possible with the limited statistics in previous observations.
Figure 6 shows a full-resolution raw 0.5-2.0~keV ACIS image of the central region
of NGC 5044.
Obvious in Figure 6 is a pair of bipolar cavities aligned along a SE-NW direction.
The significance of the inner cavities can be estimated by comparing the
counts within the cavities with that expected based on the average surface
brightness within an annulus that encloses the two cavities.  For the
NW cavity, the reduction in counts is significant at 4.3$\sigma$, while
the reduction in counts in the SE cavity is significant at 4.1$\sigma$.
The orientation of the inner pair of bipolar cavities is
consistent with the orientation of the central regions of the larger SE and NW cavities,
indicating that both AGN outbursts inflated cavities along the same axis.

\begin{figure}[t]
\center{\includegraphics[width=1.00\linewidth,bb=135 131 521 653,clip]{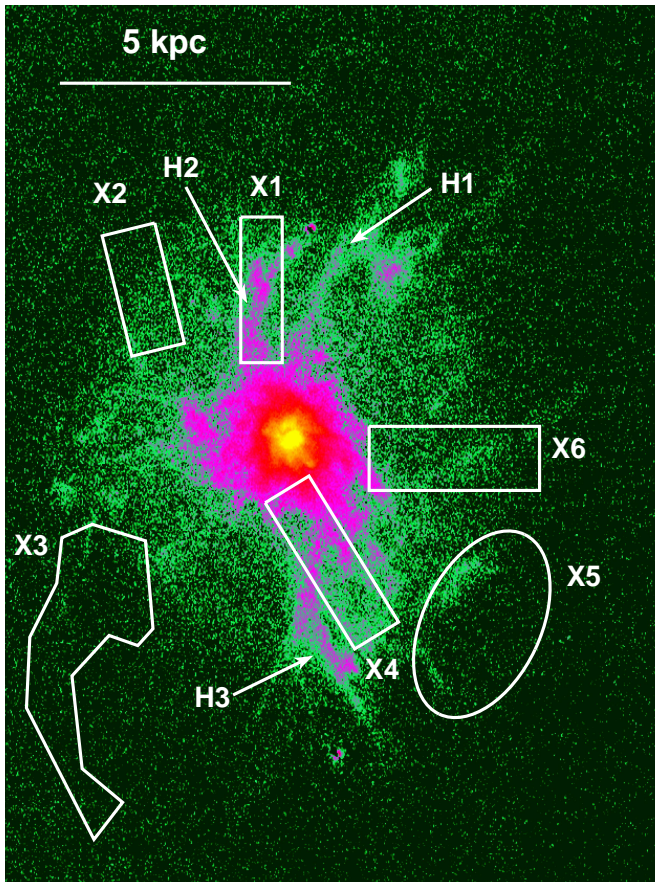}}
\caption{Comparison of the H$\alpha$ emission and the location of
the X-ray filaments identified in Figure 2. Also shown are the locations 
of the three brightest H$\alpha$ filaments.}
\end{figure}

Our ALMA observation of NGC 5044 showed that the central AGN is a bright continuum source
at 230~GHz with a flux density of 55~mJy (David et al. 2014). Since the 230~GHz flux
density exceeds the flux density of the central point source at 1.4~GHz
(Giacintucci et al. 2011),
NGC 5044 may be a less powerful version of GHz-peaked sources (O'Dea 1998), which are thought
to be radio galaxies in the early stages of an outburst.  While the inner pair of
bipolar cavities is probably still being powered by the GHz-peaked AGN, the
larger and older SE and NW cavities are no longer being powered and are buoyantly rising
within the group potential.  The complex large-scale morphology of the southern
radio lobe (Giacintucci et al. 2011) probably arises from the sloshing motion of NGC 5044
within the group potential (O'Sullivan et al. 2014).
Since the inner cavities are still being powered by the
AGN, the age of the most recent AGN outburst can be estimated from the sound crossing time,
which gives $t_{age} \sim 1$~Myr.  Assuming the southern radio lobe is
buoyantly rising at its terminal velocity, we obtain $t_{age} \sim 13$~Myr for the
older AGN outburst.

\begin{figure}[t]
\center{\includegraphics[width=1.00\linewidth,bb=124 206 477 585,clip]{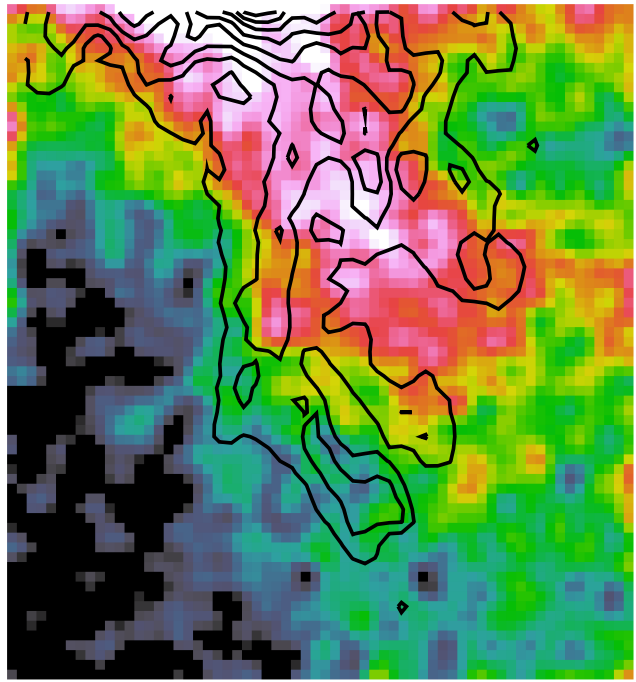}}
\caption{High-resolution comparison showing the offset between the
H$\alpha$ emission in the H3 filament (contours) and the 0.5-2.0~keV
emission in the X-ray filament X4.}
\end{figure}

There is an obvious correlation between the central dust lanes in NGC 5044
(as seen in an HST extinction image), the molecular gas, and the alignment of the
inner pair of bipolar cavities (see Figure 7). Temi et al. (2007) suggested
that the extended dust emission detected by {\it Spitzer} in NGC 5044
resulted from the disruption of a central dusty disk or torus by a recent AGN outburst.
The close agreement in alignment between the dust, molecular gas, and X-ray cavities
seen in Figure 7 certainly supports this hypothesis. Moreover, there is no kinematical
evidence in the ALMA data for a residual molecular disk.  However, it is difficult
to understand how the inner pair of bipolar cavities could have a dynamical impact
on the molecular gas, since the inner cavities cannot dredge up more mass than they
displaced, which is approximately $2 \times 10^6 M_{\odot}$.  When we assume
a galactic conversion factor between $H_2$ and CO
($X_{CO}= 2 \times 10^{20}$~cm$^{-2}$~(K km s$^{-1}$)$^{-1}$),
the total mass of the molecular gas within the inner cavities is
$2.6\times 10^7 M_{\odot}$, a factor of ten greater than the hot gas mass
displaced by the cavities.  The dust mass within the filamentary structures
seen in Figure 7 is $\sim 3 \times 10^4 M_{\odot}$ (Temi et al. 2007), which gives an unusually
high gas-to-dust mass ratio of 1000 assuming a galactic $X_{CO}$.  The $X_{CO}$
conversion factor is sensitive to many environmental factors (e.g., Bolatto et al. 2013).
Assuming a $X_{CO}$ factor ten times less than the galactic value produces
a molecular mass consistent with the mass displaced by the cavities and
a more reasonable gas-to-dust mass ratio of 100.  Thus, the alignment of
the molecular gas and cavities gives at least indirect evidence for a
lower value of $X_{CO}$ in NGC 5044.

\begin{figure}[t]
\center{\includegraphics[width=1.00\linewidth,bb=84 180 485 577,clip]{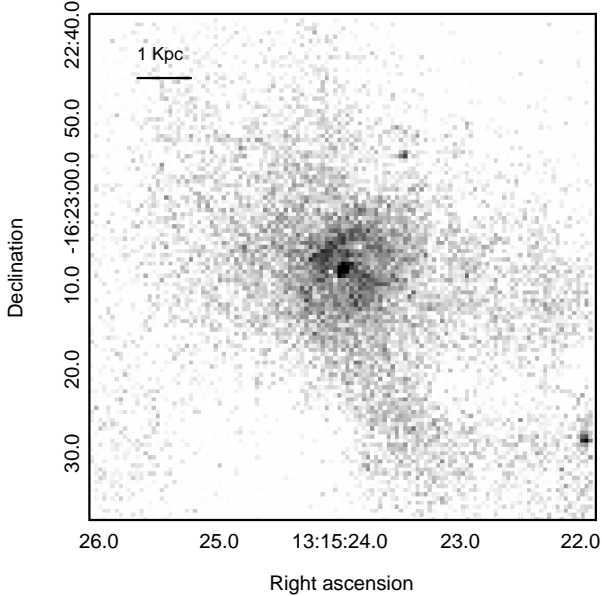}}
\caption{Raw 0.5-2.0~keV image of the combined 395~ks of {\it Chandra} ACIS data 
at full resolution ($0^{\prime\prime}.5$ per pixel.)}
\end{figure}

\section{Global X-Ray Properties}

To fully investigate the thermodynamic state of the X-ray filaments,
we need to determine a number of basic properties
of the NGC 5044 group, including the gas pressure profile, total mass distribution,
gravitational potential, free-fall time, and radiative cooling time.  All of these
quantities depend on derivations of the gas density and temperature distributions,
which is the focus of this section.  While azimuthally averaged gas density and temperature
profiles have been derived previously from archival {\it chandra} data for NGC 5044, we currently
have four times the data at our disposal, so it is worth repeating the analysis.

The azimuthally averaged deprojected temperature profile of NGC 5044 is
shown in Figure 8.  The temperature profile
was derived by extracting spectra from all seven ObsIDs in concentric annuli
with widths ranging from $2^{\prime\prime}$ at small radii to
$10^{\prime\prime}$ at large radii. The central $2^{\prime\prime}$ is
excluded from the analysis because of the X-ray emission from the central AGN.
The seven spectra extracted from each annulus were then simultaneously fit to an absorbed
single-temperature model {\it phabs*apec} in XSPEC with the hydrogen column density
fixed to the galactic value and the temperature and abundance of heavy elements
treated as free parameters.  The XSPEC tool {\it projct} was used to deproject the spectra.
Figure 8 shows that the deprojected temperature
profile is well fit with an analytic function given by

\begin{figure}[t]
\center{\includegraphics[width=1.00\linewidth,bb=64 190 470 590,clip]{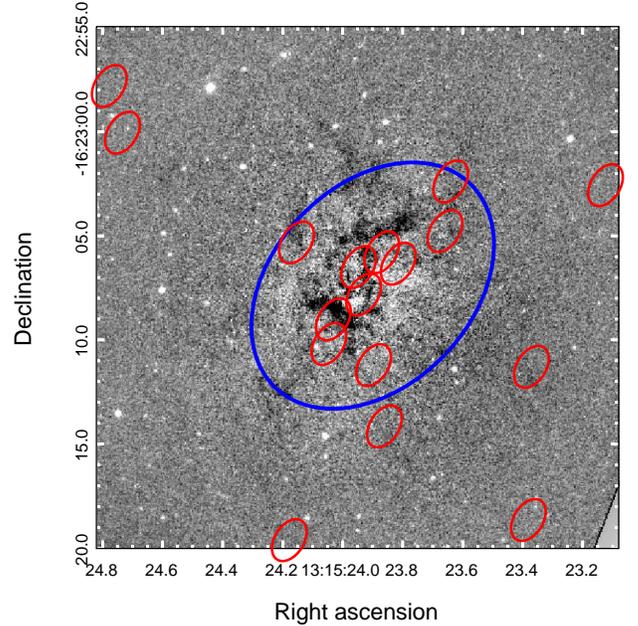}}
\caption{Comparison of the dust lanes visible in an HST extinction image, extent of
the inner pair of X-ray cavities (cyan ellipse), and molecular structures detected with ALMA 
(red ellipses with the size of the ALMA cycle 0 beam).}
\end{figure}

\begin{equation}
T(r) = T_0 \left( { {r} \over {r_c} } \right)^{-p} + T_{min},
\end{equation}

\noindent
which is similar to the temperature profile used by Allen et al. (2001) for
the inner region of cooling flows. The best-fit parameters for NGC 5044 are
$T_0=0.166 \pm 0.005$, $p=0.846 \pm 0.023$, and $T_{min}=0.782 \pm 0.005$, 
with $r_c$ fixed at 10.0~kpc.
The largest deviation between the data and analytic model occurs around
4~kpc from the central AGN, which coincides with the location of the
SE cavity.

The azimuthally averaged density profile was computed from the
best-fit emission measures derived in the deprojected spectral analysis
(see Figure 9). We first tried to fit the density profile with
an analytic function similar to that used by Sun et al. (2009)
and Vikhlinin et al. (2006), but found that an extra power-law component
was required to account for the emission within the central few kpcs.
Fitting the density profile to a model given by

\begin{equation}
n_e(r) = a_0 \left( { {r} \over {r_0} } \right)^{-\alpha} 
+ a_1 \left( { {r} \over {r_c} } \right)^{-\gamma}\left[ 1 + \left( { {r} \over {r_c} } \right)^2 \right]^{-3\beta/2+\gamma/2},
\end{equation}

\noindent
we derived best-fit values of $a_0=0.0314 \pm 0.0128$, $\alpha=1.80 \pm 0.32$,
$a_1=0.0113 \pm 0.0034$,
$r_c=21.1 \pm 6.6$, $\gamma=0.394 \pm 0.148$, and $\beta=0.790 \pm 0.144$, with $r_0$ fixed at 1~kpc. Figure 9 shows that $n_e \sim r^{-1.2}$ between 0.3 and 1.0~kpc.
To determine if the density profile at smaller radii
follows the same power-law profile, we extracted a spectrum from within the
central $2^{\prime\prime}$ region and fit the spectrum to an absorbed power-law
(to account for the emission from the central AGN) plus thermal model.  Assuming a
power-law for the density profile ($n_e \sim r^{-p}$), with p=0, p=1/2, and p=1,
we derived the allowed density profiles that would reproduce the observed
emission measure (see Figure. 9).  This calculation shows that the
density profile must flatten within the central 300~pc in order to not exceed
the observed emission measure within this region.

\begin{figure}[t]
\center{\includegraphics*[width=1.00\linewidth,bb=18 147 572 702,clip]{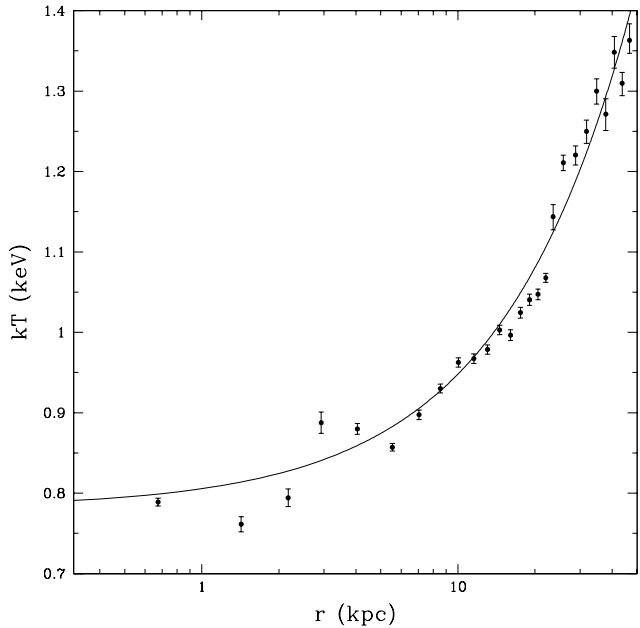}}
\caption{Azimuthally-averaged, deprojected temperature profile along with the best-fit
analytic model given in eq. (1).} 
\end{figure}

\section{Where Does the Gas Cool, Exactly?}

Many studies, both observational and theoretical, have investigated
the criteria that must be met for the central gas in clusters
to cool out of hot phase.
Observations show that multiphase gas (primarily identified by H$\alpha$
emission) and recent star formation are only present in systems with
central cooling times shorter than $5 \times 10^8$~years (Rafferty et al. 2008)
or central entropies lower than 30~keV~cm$^{2}$ (Cavagnolo et al. 2008,2009).
Voit et al. (2008) showed that these two criteria both result from the
competing effects of radiative cooling and thermal conduction.
Conduction is unable to suppress radiative cooling in systems with
lower entropies based on the Field criterion. In addition, Werner et al. (2014)
used the Field criterion to show that [CII] emission, which originates
from the photodissociation region surrounding molecular clouds,
arises only from groups, including NGC 5044, that have thermally unstable
gas over an extended region.

Numerical simulations (McCourt et al. 2012; Sharma et al. 2012;
Gaspari et al. 2012,2013,2015; Li \& Bryan 2014a,2014b) suggest that
a separate criterion that governs the development of a multiphase
medium in clusters.  These simulations find that perturbations in a globally
stable atmosphere (i.e., a balance between heating and cooling over large
spatial and temporal scales) only become nonlinear if $t_c/t_{ff} < 10$,
where $t_c$ is the isochoric cooling time and $t_{ff}$ is the free-fall
time\footnote{Some authors use the thermal instability timescale, $t_{TI}$, instead of
$t_c$. Using the equations in Sharma et al. (2012),
we obtain $t_{TI}=0.8 t_c$ for the average gas temperature in NGC 5044.}
These studies suggest that the
AGN is fueled by the accretion of cold gas, as in Pizzolato \& Soker (2005,2010),
and not by Bondi accretion of hot gas.

\begin{figure}[t]
\center{\includegraphics*[width=1.00\linewidth,bb=18 157 567 694,clip]{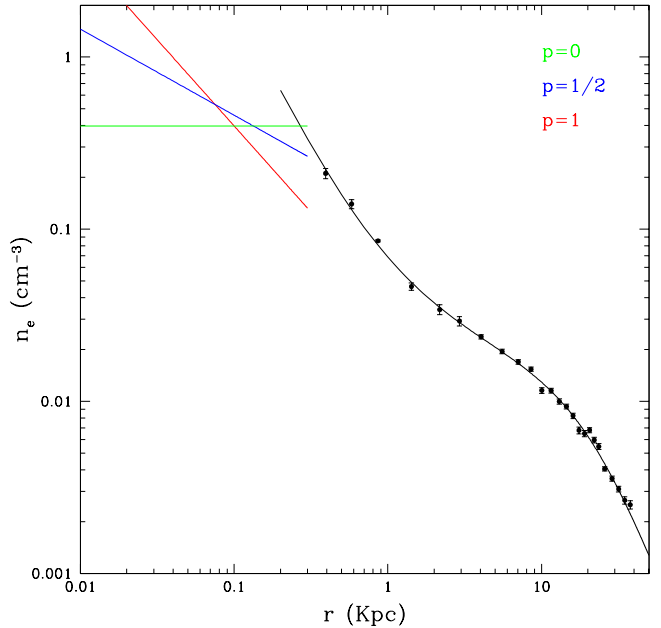}}
\caption{ Azimuthally-averaged electron number density profile derived from the deprojected
spectral analysis described in the text along with the best-fit analytic model shown
in eq. (2). The solid lines show the density profiles required to reproduce the 
observed emission measure with the central $2^{\prime\prime}$ assuming a
power-law profile with three different indices.}
\end{figure}

Based on a subsample of ACCEPT clusters with H$_{\alpha}$ measurements,
Voit \& Donahue (2015a) found that the entropy and $t_c/t_{ff}$
criteria serve as complementary indicators for the presence
of multiphase gas in clusters.  In a subsequent paper, Voit et al.
(2015b) developed a ``precipitation" model for the growth of galaxies
based on the $t_c/t_{ff}$ criterion.  However, McNamara et al. (2016)
recently claimed that $t_c$ alone was a more reliable indicator for the
presence of H$\alpha$ emission and star formation than the $t_c/t_{ff}$
ratio. The presence of some clusters with short cooling times,
but no indications of cooler gas or star formation (e.g., A2029),
suggests that an additional process must be present in
the cores of clusters to trigger thermal instability.
McNamara et al. thus proposed a ``stimulated feedback" mechanism
that requires the presence of buoyantly rising radio cavities
that lift up low-entropy gas as an essential ingredient in the
production of molecular gas and star formation. Along the same
lines of argument as McNamara et al., Brighenti et al. (2015)
presented numerical simulations of NGC 5044 and found that buoyantly rising
cavities trigger episodes of enhanced cooling.

Using the deprojected spectroscopic results, we computed $t_c/t_{ff}$
within the central 40~kpc of NGC 5044 (see Figure 10).
We determined the total gravitating mass and free-fall time
($t_{ff}=\sqrt{2r/g}$, where $g$ is the local acceleration of gravity)
assuming hydrostatic equilibrium, and Equations (1) and (2)
for the temperature and density profiles of the hot gas.
In addition, we also computed the free-fall time assuming an
isothermal sphere and the stellar velocity dispersion for NGC 5044,
where $g=2 \sigma_*^2/r$, as done in other studies (e.g., Gastaldello et
al 2009; Voit \& Donahue 2015).
The H$_{\alpha}$ filaments in NGC 5044 extend to approximately 8~kpc from the
central AGN, mostly in a north-south direction (see Figure 4).  Werner et al.
(2014) found that the [CII] emission also has a similar extent.
Figure 10 shows that the observed 8~kpc extent of the H$\alpha$
and [CII] emission corresponds to the radius where $t_c/t_{ff}$ attains its minimum
value of approximately 12. More sensitive observations could
potentially detect more extended H$\alpha$ and [CII] emission, but $t_c/t_{ff}$ remains
lower than 30 within the central 25~kpc, which is three times the observed
extent of the H$\alpha$ and [CII] emission.
Within the central 8~kpc in NGC 5044, the radiative cooling time of the hot
gas is shorter than 500~Myr and the entropy, $K=T/n_e^{2/3}$, is lower than 15~keV~cm$^{-2}$.
Since NGC 5044 also hosts two pairs of bipolar X-ray cavities, NGC 5044
satisfies all suggested criteria for the presence of multiphase gas.
Below we identify the exact location of the thermally unstable multiphase
gas.

\begin{figure}[t]
\center{\includegraphics*[width=1.00\linewidth,bb=19 144 567 696,clip]{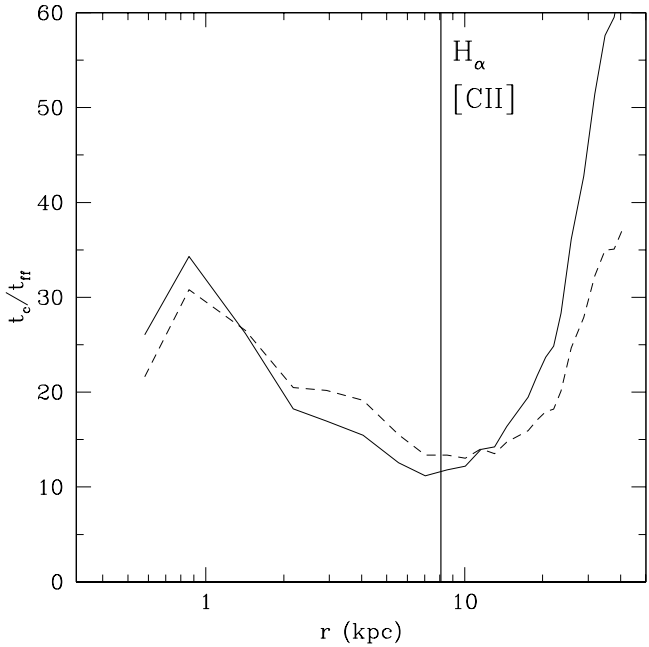}}
\caption{Ratio of the cooling time ($t_c$) of the 
hot gas and the free-fall time ($t_{ff}$) within the central region
of NGC 5044.  The free-fall time is computed in two ways:
the assumption of hydrostatic equilibrium (solid line) and
the assumption of an isothermal sphere (dashed line).
The vertical line indicates
the extent of the H$_{\alpha}$ and [CII] emission.} 
\end{figure}

\subsection{Deprojected Two-Temperature Results}

While the H$_{\alpha}$ and [CII] emissions reveal the presence of $10^4$~K
and 100~K gas within the central 8~kpc of NGC 5044, a key question is
within which radius the X-ray emitting gas is actually cooling.
To address this question, we extracted spectra from all seven ObsIDs
in concentric annuli beyond $2^{\prime\prime}$ from the central AGN with widths of
$10^{\prime\prime}$ and fit these spectra simultaneously
with a deprojected two-temperature thermal {\it apec} model with the
abundances linked between the two thermal components and the hydrogen
column density fixed to the galactic value.  The results of the
deprojected spectral analysis are shown in Table 2.  For spectra
extracted beyond $52^{\prime\prime}$ (7.8~kpc), the addition of a second
temperature component did not produce a
statistically significant improvement in the fit above that attained with a
single-temperature model.  This radius is remarkable similar to the observed
extent of the H$_{\alpha}$ and [CII] emission and the radius where $t_c/t_{ff}$
attains its minimum value.  Assuming pressure balance
between the two thermal components, the volume-filling factor of the cooler
component is

\begin{equation}
f_{vol} = \left[ 1 + \left( { T_h \over T_c} \right)^2 \left( { \epsilon_h \over \epsilon_c} \right) \right]^{-1},
\end{equation}

\noindent
where $\epsilon_h$ and $\epsilon_c$ are the emission measures of the
hot and cool components, respectively.  For the outer spectra, we fixed
the temperature of the cooler component at 0.5~keV to derive 90\% upper limits
on the emission measure of the cooler thermal component.  Figure 11 shows a 
a sharp transition around 8~kpc between the outer regions, which
have $f_{vol}<0.01$, and the inner regions, which have $f_{vol}=0.03-0.1$.
Table 2 shows that the temperature ratio between the two thermal
components is roughly a factor of two, indicating that the mass fraction
of the cool component increases up to approximately 20\% within the central 1~kpc.
This study thus confirms that not only
is there multiphase gas in the form of H$_{\alpha}$ and [CII] emission
within the central 8~kpc, but also that this is the region where the
X-ray emitting gas is actually cooling out of the hot phase.

\begin{figure}
\center{\includegraphics[width=1.00\linewidth,bb=19 146 576 699]{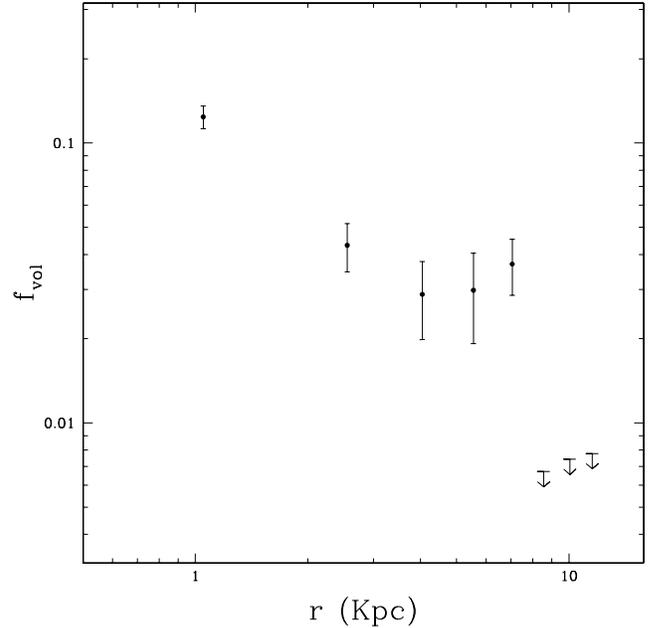}}
\caption{Volume-filling factor of cooler gas, $f_{vol}$, derived from a 
two-temperature deprojection of the ACIS data.} 
\end{figure}

\begin{deluxetable*}{p{2.5cm}cccc}
  \tablecaption{Deprojected Two-Temperature Results}
  \tablehead{
    \colhead{Radii} &
    \colhead{$\rm{kT_c}$} &
    \colhead{$\epsilon_c$} &
    \colhead{$\rm{kT_h}$} &
    \colhead{$\epsilon_h$}
    \\
    \colhead{(arcseconds)}  &
    \colhead{(keV)} & 
    \colhead{($10^{-5}$~cm$^{-5}$)} &
    \colhead{(keV)} &
    \colhead{($10^{-5}$~cm$^{-5}$)}
  }
  \startdata
  ~~~~~~~~~~~2-12  & 0.530 (0.483 - 0.549) & 14.6 (13.6 - 16.4) & 0.862 (0.851 - 0.864) & 39.1 (37.5 - 40.9) \\
  ~~~~~~~~~~12-22 & 0.434 (0.422 - 0.474) & 3.43 (2.82 - 4.19) & 0.861 (0.857 - 0.865) & 19.3 (18.3 - 20.3) \\
  ~~~~~~~~~~22-32 & 0.439 (0.422 - 0.544) & 4.09 (3.10 - 5.71) & 0.864 (0.858 - 0.874) & 35.5 (34.3 - 36.3) \\
  ~~~~~~~~~~32-42 & 0.536 (0.391 - 0.572) & 3.82 (2.69 - 5.49) & 0.862 (0.858 - 0.864) & 48.2 (47.4 - 49.6) \\
  ~~~~~~~~~~42-52 & 0.433 (0.423 - 0.448) & 9.51 (8.12 - 12.6) & 0.873 (0.863 - 0.879) & 60.8 (59.8 - 61.3) \\
  ~~~~~~~~~~52-62 & 0.5                   & $<1.94$            & 0.881 (0.875 - 0.889) & 92.2 (90.2 - 93.5) \\
  ~~~~~~~~~~62-72 & 0.5                   & $<1.52$            & 0.948 (0.945 - 0.954) & 58.0 (56.9 - 59.1) \\
  ~~~~~~~~~~72-82 & 0.5                   & $<2.20$            & 0.969 (0.964 - 0.975) & 73.7 (72.0 - 74.7) \\
  \enddata
\end{deluxetable*}

\subsection{A Method for Detecting Multiphase Gas}

Multi-phase gas can be detected from the width of the blended Fe-L lines, which
requires far fewer photons than a full spectral analysis.  Most of the X-ray emission
from 1~keV groups arises from L-shell transitions
from Fe with ionization states between Fe XIX (Ne-like) and Fe XXIV (He-like).
With CCD spectral resolution, these L-shell lines are blended into a single
feature between approximately 0.7 and 1.2~keV. To calibrate the effect of
multiphase gas on the width of the blended Fe-L line, we generated a grid
of simulated single- and two-temperature thermal models
with a range of temperatures, hydrogen column densities, and abundance of heavy elements.
For each simulated spectrum, we computed the 25th, 50th and 75th percentiles in the
cumulative photon number distribution between 0.7 and 1.2~keV.  Figure 12
shows how the width of the blended Fe-L lines (the difference between the
75th and 25th percentile energies) varies with the median or 50th percentile energy
(which is a measure of the emission-weighted temperature) for each simulated spectrum.
While there is some trend of increasing Fe-L linewidth with decreasing temperature, all
single-temperature models have have fairly narrow linewidths. Varying the abundance
of heavy elements and N$_H$ also has little effect on the Fe-L linewidth in
single-temperature spectra.  For the simulated two-temperature models, we set the
temperature of the cooler component to one-half of the temperature of the hotter
component and fixed the emission measures for the two components
so that each thermal component produced the same 0.5-2.0~keV flux.  Figure 12 shows
that the simulated two-temperature spectra have significantly broader Fe-L
linewidths that the single-temperature models for a given median photon energy.

To compare the results of the simulated spectra with the NGC 5044 data, we
adaptively binned the ACIS data to a S/N of 30 within the 0.7-1.2~keV energy band and
computed the 25th, 50th, and 75th percentile photon energies within each binned region.
The resulting values are plotted in Figure 12 and fully span the region between
the simulated single- and two-temperature results. Regions containing the
highest level of multiphase gas (i.e., Fe-L linewidths greater than 190 eV)
are shown in color in Figure 13.  Regions with narrower Fe-L linewidths
are shown in black in this figure.  As seen in Figure 13, the regions with
the broadest Fe-L linewidths coincide with the X4 and X5 filaments,
and the regions with the narrowest Fe-L linewidths
fill the regions between the filaments. Actually, the region containing
the greatest degree of multiphase gas lies to the SE of the X4 filament and
is coincident with the H3 filament (see Figure 5).

Of interest in Figure 13 is an additional region toward the NW (labeled T1)
with a broad Fe-L linewidth.  This region is unremarkable in the X-ray image,
but lies in the same direction as the NW cavity.  Region T1
is well outside of the detected H$\alpha$ and [CII] emission and
approximately twice as far from the central AGN as the other multiphase filaments.
Since region T1 lies along the same direction as the NW cavity, it may contain
low-entropy gas that was dredged up by an earlier AGN outburst.  A more detailed
spectroscopic analysis of the identified multiphase regions is presented below.

\vspace{0.5in}

\begin{figure}
\center{\includegraphics[width=1.00\linewidth,bb=20 152 573 701,clip]{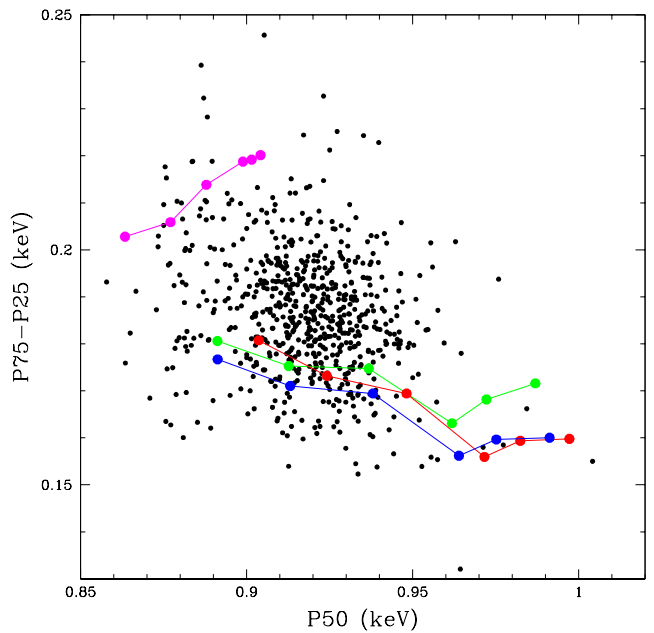}}
\caption{Curves with points show the correlation between the 
Fe-L linewidth (P75-P25) and the median photon energy (P50) based on a grid 
of simulated single- and two-temperature thermal spectra, where P25, P50, and P75 
are defined in $\S~5.3$.  The lower three curves show the results of
single-temperature thermal models with temperatures ranging from 
0.8 (leftmost point) to 1.3~keV (rightmost point).  The green curve
corresponds to $\rm{N}_{H}=\rm{N}_{gal}$, and an abundance of 
heavy elements, Z=0.6, the red curve to $\rm{N}_{H}=2\rm{N}_{gal}$
and Z=0.6, and the blue curve to $\rm{N}_{H}=\rm{N}_{gal}$ and Z=1.0. 
The upper red curve shows
the results of a set of simulated two-temperature thermal models with
$\rm{N}_{H}=\rm{N}_{gal}$ and Z=0.6, the same temperature range for the hotter 
temperature component, and a temperature for the cooler component equal 
to one-half of the temperature of the hotter component.  Also shown are the 
values from the adaptively binned 0.7-1.2~keV 
ACIS image of NGC 5044.}
\end{figure}

\begin{figure}
\center{\includegraphics[width=1.00\linewidth,bb=38 120 576 649,clip]{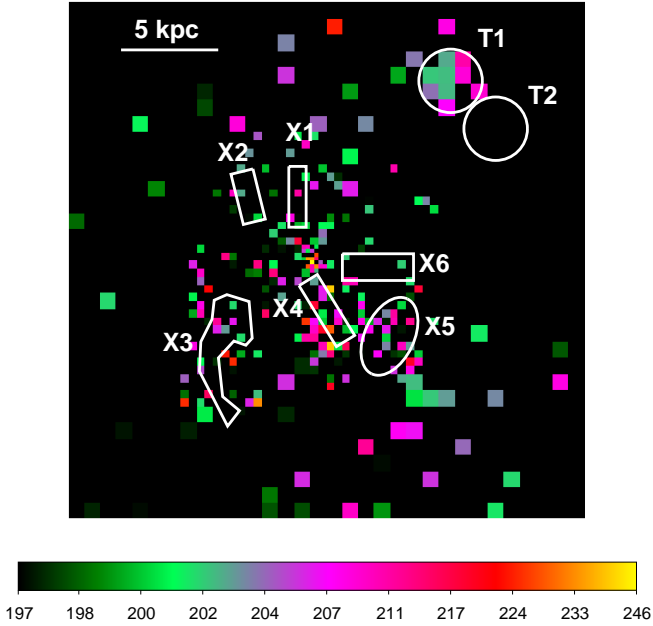}}
\caption{Fe-L linewidth image. Only regions with a Fe-L linewidth exceeding 
190~eV are shown in color. All other regions are shown in black. The labelled 
regions are the same as those in Figure 2, except for the addition of 
regions T1 and T2.}
\end{figure}

\subsection{Multiphase Gas in the X-ray and H$\alpha$ Filaments}

The azimuthally averaged spectroscopic analysis of the ACIS data only
shows evidence for multiphase within the central 8~kpc, while the Fe-L linewidth
method also detects an additional region at larger radii with multiphase gas.
In this section, we perform a detailed spectroscopic analysis of the X-ray
filaments and the additional region with a large Fe-L linewidth to determine
the exact location of cooling gas and the underlying conditions in these regions.
ACIS spectra were extracted from all regions shown in Figures 2 and 13
and fit to absorbed single- and two- temperature thermal models
with the hydrogen column density fixed at the galactic value.
Other models were also fit to the spectra (e.g., a partial
covering model, a variable elemental abundance model, and a charge-exchange model),
but none of these models produced a statistically significant improvement
in the fit above that attained with a two-temperature model.
While the temperature of the gas in the filaments only varies by about 20\%
(see Table 3), it is clear that the X4 filament is the coolest and
highest X-ray surface brightness filament in NGC 5044.
We also note that region T1 (broad Fe-L linewidth) is about 10\% cooler
that the nearby region T2 (narrow Fe-L linewidth) used for comparison.
In Table 4 we show the results of fitting a two-temperature thermal
model to all the regions shown in Table 3. Since the F-test cannot
be used to compare the single- and two-temperature results
(Protassov et al. 2002), we generated 1000 simulations of the best-fit
single-temperature model for each region shown in Table 3, fit
the simulated spectra to single- and two-temperature models,
and computed the cumulative distribution of $\Delta \chi^2$
between the single- and two-temperature fits. Only regions
with a $\Delta \chi^2$ greater than 95\% of the simulations
are listed in Table 4.  This analysis
shows that the X4 and X5 filaments and the T1 region have the
greatest fraction of multiphase gas, which is consistent with
our Fe-L linewidth diagnostic (see Figure 13).

To illustrate the differences between the filaments, we coadded the
spectra extracted from the observations between 2015 May and August,
during which time there was little change in the ACIS response.
A comparison of the coadded spectra for the T1 and T2 regions is shown
in Figure 14.  Even though these regions are adjacent to one another
and show no morphological features, there is clear evidence for a
soft excess in region T1.  Figure 14 also shows that it is not the
case that the gas in region T1 is cooler than the gas in region T2.
Above 1~keV, the spectra are essentially the same, indicating
that both regions contain ambient gas of the same temperature, but
only region T1 contains an additional component of cooler gas.

Of particular interest is a comparison between the X6 filament (no evidence for a
cooler thermal component and little association with H$_{\alpha}$ emission)
and the X4 filament (strong evidence for cooler gas and significant H$_{\alpha}$
emission).  These two filaments have very similar X-ray morphologies
(see Figure 2).  A comparison of the coadded spectra from the 2015 observations
for the two filaments is shown in Figure 15 after renormalizing the spectra
to yield the same 1.5-3.0~keV count rate.  Similar to the comparison between
regions T1 and T2, the shape of the spectra is nearly identical above 1~keV,
but only the X4 filament clearly shows a soft excess, indicative of
cooling gas.  Using the emission measure from the best-fit single-temperature
model and assuming a cylindrical geometry for the X4 and X6 filaments,
we find that the cooling times in the X4 and X6 filaments are essentially
the same ($1.8 \times 10^8$~years for
the X4 filament and $1.9 \times 10^8$~years for the X6 filament). Since the two
filaments reside at approximately the same distance from the central AGN, the
free-fall times are also very similar.  Thus, both X-ray filaments have very
similar cooling times and free-fall times, but only one contains cooling gas
and H$_{\alpha}$ emission.

\begin{deluxetable}{ccccc}
 \tablecaption{Single-Temperature Fit to Selected Regions}
 \tablehead{
 \colhead{Region} &
 \colhead{kT} &
 \colhead{Z} &
 \colhead{$\chi^2$} &
 \colhead{DOF}
 \\
 \colhead{} &
 \colhead{(keV)} &
 \colhead{(Z$_{\odot}$)} &
 \colhead{} &
 \colhead{}
 }
 \startdata
X1 & 0.831 (0.823-0.839) & 0.53 (0.48-0.58) & 268.5 & 234 \\
X2 & 0.903 (0.895-0.912) & 0.64 (0.58-0.72) & 306.2 & 247 \\
X3 & 0.939 (0.934-0.944) & 0.71 (0.67-0.76) & 544.4 & 440 \\
X4 & 0.756 (0.750-0.761) & 0.49 (0.46-0.53) & 447.5 & 364 \\
X5 & 0.823 (0.819-0.828) & 0.60 (0.57-0.64) & 497.3 & 447 \\
X6 & 0.909 (0.903-0.915) & 0.79 (0.73-0.86) & 501.8 & 393 \\
\hline
T1 & 1.02 (1.01-1.03)    & 0.48 (0.44-0.53) & 240.0 & 192 \\
T2 & 1.14 (1.12-1.16)    & 0.70 (0.61-0.80) & 219.6 & 169 \\
\enddata
\end{deluxetable}

\begin{deluxetable}{ccccc}
 \tablecaption{Two-Temperature Fit to Selected Regions}
 \tablehead{
 \colhead{Region} &
 \colhead{kT$_c$} &
 \colhead{kT$_h$} &
 \colhead{$\chi^2$} &
 \colhead{DOF}
 \\
 \colhead{} &
 \colhead{(keV)} &
 \colhead{(keV)} &
 \colhead{} &
 \colhead{}
 }
 \startdata
X4 & 0.543 (0.456-0.601) & 0.776 (0.768-0.783) & 437.2 & 363 \\
X5 & 0.432 (0.395-0.514) & 0.812 (0.807-0.819) & 477.9 & 446 \\
T1 & 0.685 (0.620-0.768) & 1.06 (1.05-1.09)    & 227.1 & 191 \\
\enddata
\end{deluxetable}

One possible explanation for the differences between the X4 and X6 filaments is
that filament X4 contains lower entropy gas that was dredged up by
a recent AGN outburst, while filament X6 is composed of gas that is still
falling inward. 
If filament X4 contains dredged-up gas,
the abundance of heavy elements in this filament should be enhanced
as a result of enrichment from supernovae, while the gas resides within the 
central galaxy.
However, a simple examination of Figure 15 shows that the Si abundance (which
is produced by both SNe~Ia and SNe~II) is very similar between the two filaments.
A full spectral analysis also reaches the same conclusion.  A comparison
of the Fe abundance (which is primarily produced by SNe~Ia) would be more
informative, but it is very difficult to derive tight constraints
on the Fe abundance in multiphase gas with CCD resolution spectra (Buote 1999).

\section{Kinematics of Molecular Gas}

The observed velocities of the giant molecular associations (GMAs)
detected by ALMA can be used to determine their place of origin, assuming the
clouds are able to fall unimpeded under the influence of gravity.
We computed the gravitational
potential of NGC 5044 directly from the best-fit density and temperature distributions
given in eqs. (1) and (2) and the assumption of hydrostatic equilibrium.  Figure 16
shows the distance over which each of the 24 GMAs must fall radially inward
to achieve its observed line-of-sight velocity for a range of inclination angles
(i.e., the orientation of the cloud trajectory relative to the plane of the sky
with $i=0$ corresponding to motion within the plane of the sky).
The required infall distance for a given line-of-sight velocity increases
with decreasing inclination angle, since the inferred space velocity increases
with decreasing inclination angle.  Figure 16 shows that the GMAs can attain their
observed velocities through unimpeded gravitational infall over distances of
only 0.1-1~kpc.  Since we detect multiphase gas out to 15~kpc, unimpeded
infall of thermally unstable clouds is essentially ruled out.
If the kinematics of the GMAs are gravitational in origin, then
the cooling clouds must be pinned to the hot gas, possibly
by magnetic fields, or experience significant drag during most of the
cooling process.

\begin{figure}[t]
\center{\includegraphics[width=1.00\linewidth]{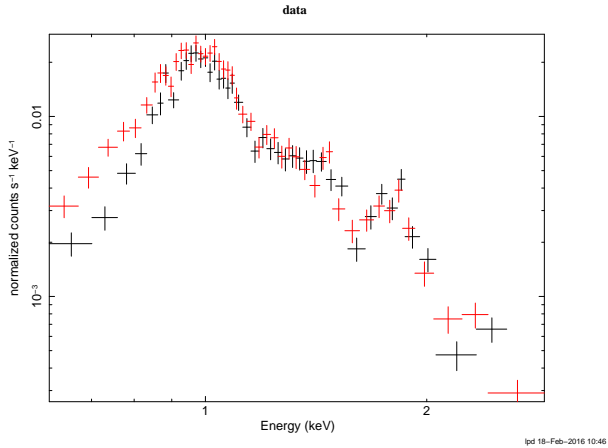}}
\caption{A comparison of the coadded ACIS spectra for regions T1 (red) 
and T2 (black).} 
\end{figure}

It is interesting to compare our results with a study of the molecular gas
in PKS 0745-191 by Russell et al. (2016).  Even though NGC 5044 is a group
and PKS 0745-191 is a massive cluster, the central dominant galaxies
in these two systems have very similar gravitating masses of
$\sim 2 \times 10^{14} M_{\odot}$ within 30~kpc (David et al. 2009;
Sanders et al. 2014). However, PKS 0745-191 has clear X-ray cavities
exterior to the molecular gas, while NGC 5044 shows
no evidence for X-ray cavities beyond the ends of X-ray and H$\alpha$ filaments.
In addition, the cavity power in PKS0724-191 is a factor of 1000 times
greater than that in NGC 5044 (David et al. 2009).  The ALMA observation of
the central galaxy in PKS 0745-191 reveals three molecular
filaments with systemic velocities lower than 100~km~s$^{-1}$ and FWHMs for
the CO(1-0) and CO(3-2) lines smaller than 150~km~s$^{-1}$ (Russell et al. 2016).
For comparison, the FWHM of the CO(2-1) line derived from the IRAM 30m
observation of NGC 5044 is 220~km~s$^{-1}$ (David et al. 2014).
Based on the observed velocities, Russell et al. conclude, as we do for NGC 5044, that
the velocities of the molecular structures are too low to be produced by
uninhibited gravitational infall.  They suggest that the
molecular gas arises from in situ cooling of warm gas dredged up by
the AGN-inflated cavities.  While we both arrive at similar conclusions regarding the
influence of gravity on the kinematics of the molecular gas, it is
interesting that the FWHM of the CO line in NGC 5044 is broader
than that in PKS0724-191, even though the cavity power in NGC 5044
is a factor of 1000 times lower than the cavity power in PKS0724-191.

There is no evidence for an extended component of CO emission
associated with the X-ray or radio cavities in NGC 5044, unlike the
case in the Phoenix cluster (z=0.596), where the CO emission
drapes the two large X-ray cavities (Russell et al. 2017).
This could simply be due to the large contrast in spatial
scales that are sampled by the ALMA data.  Interferometric observations
are only sensitive to emission over a limited range of angular scales.
Our cycle 0 ALMA data had a beam size of $2^{\prime\prime}.1$ (300~pc) and a
maximum recoverable size (MRS) of $9^{\prime\prime}$ (1.3~kpc),
while the ALMA data presented in Russell et al. (2017) had
a beam size of $0^{\prime\prime}.6$ (4~kpc).  Thus, our ALMA
data are insensitive to molecular structures on the same scales
as those detected in the Phoenix cluster. It may be that the
more extended and diffuse CO emission detected in the Phoenix cluster is molecular
gas that was recently uplifted or cooled in situ in the wake of the
X-ray cavities, as argued in Russell et al. (2017),
and the molecular structures detected by ALMA in NGC 5044
are more evolved structures that have since decoupled from
the hot gas.

\begin{figure}[t]
\center{\includegraphics[width=1.00\linewidth]{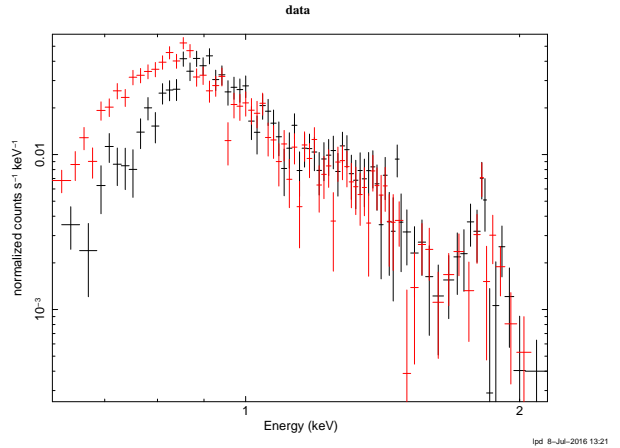}}
\caption{Comparison between the ACIS spectra of the X4 (red) and X6
(black) filaments.}
\end{figure}

Unlike the molecular gas in NGC 5044, the velocity dispersions of the
100~K gas (as seen in [CII]) and $10^4$~K gas (as seen in H~$\alpha$)
are much greater.  Hamer et al. (2016) measured a FWHM for
the H~$\alpha$ emission of 457~km~s$^{-1}$ in NGC 5044, which is a
fairly typical FWHM for the 73
clusters in their sample.  Werner et al. (2014) presented the results
of a [CII] study of six cooling flow groups using {\it Herschel} data.
While Werner et al. do not list the FWHM of the [CII] line in NGC 5044,
the [CII] velocity dispersion obtained from their velocity dispersion
map (Figure 4 in Werner et al.) is about $180$~km~s$^{-1}$, which gives
a FWHM for the [CII] line of 420~km~s$^{-1}$. Thus, both the
H~$\alpha$ and [CII] emitting gas have significantly greater
velocity dispersions than the CO emitting gas.  The broad [CII] emission
line is particularly hard to understand, since [CII] emission originates
in the photodissociation region surrounding molecular clouds.

\begin{figure}[t]
\center{\includegraphics[width=1.00\linewidth,bb=20 148 570 696,clip]{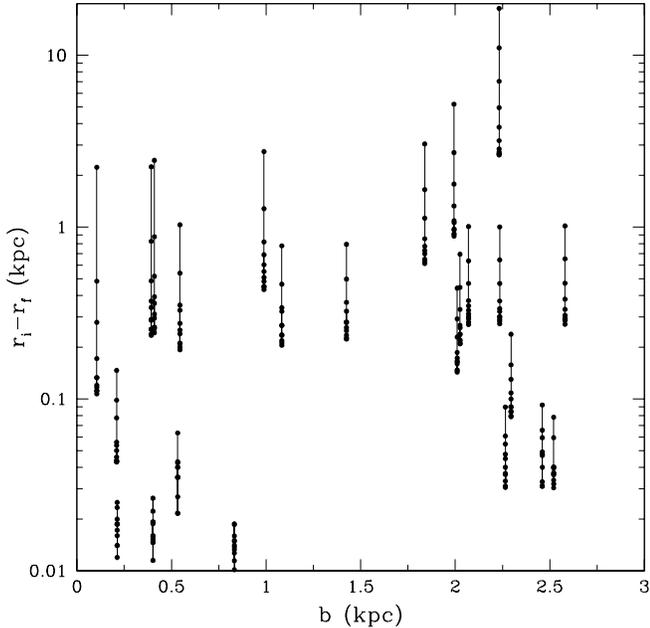}}
\caption{Length ($r_i-r_f$) over which each of the 24 GMAs detected 
by ALMA in NGC 5044 must fall radially inward to attain their observed
line-of-sight systemic velocity for a range of values of the inclination 
angle of the GMAs trajectory relative to the plane of the sky 
($i$ = 20, 30, 40, 50, 60, and 70$^{\circ}$). The required infall distances are
plotted vs. the projected distance of each GMA from the central AGN.} 
\end{figure}

\section{Discussion}

As discussed above, several criteria have been suggested for the
onset of cooling in the central region of clusters, including 
the central cooling time, the central entropy, the ratio of
the cooling time to the free-fall time, and the presence of
X-ray cavities that dredge up low-entropy gas.
The NGC 5044 group satisfies all these criteria. Since we only
present the results for a single object, we cannot
shed light on which of these criteria is the fundamental
requirement for the production of cold gas, but we can shed light
on exactly where the gas does cool and which 
conditions prevail in these regions.

An azimuthally averaged analysis of the X-ray emission from the
NGC 5044 group shows that the hot gas is thermally unstable, based on
the empirical evidence for multiphase gas, within the central 8~kpc.
This radius coincides with the extent of the H$\alpha$ and [CII] emission
and the minimum value of $t_c/t_{ff}$.  A more spatially resolved analysis
shows that only a small fraction of the gas within the central 8~kpc is
thermally unstable.  Based on the Fe-L linewidth, which is a good
diagnostic for multiphase gas, coupled with a detailed spectroscopic analysis of
selected regions, the bulk of the multiphase gas is confined to the
filamentary X-ray structures.

A spectroscopic analysis of the X-ray filaments shows that they
display a variety of properties.  The X-ray filaments X2, X3, and X6
contain gas that is cooler than the ambient gas, but there is no evidence of a
soft X-ray excess or any association with H$\alpha$
filaments.  The X-ray filament X5 shows evidence
of a soft X-ray excess but there is only some faint diffuse H$\alpha$ emission
coincident with the filament. The X-ray filament X4 has the most significant
soft X-ray excess and is also coincident with the brightest H$\alpha$
filament.  Thus, the differences between the X-ray filaments might
simply reflect the evolution of thermally unstable gas from the initial
production of a cool filament, to the development of a soft X-ray excess,
and finally to the production of lower ionization gas.  The only exception to
this scenario is filament X1, which does not show any evidence of a
soft X-ray excess, but is coincident with an H$_{\alpha}$ filament.  However,
filament X1 has the poorest photon statistics of any filament, so this could
just be a sensitivity issue.

Other than in situ cooling, the X-ray filaments could result from the
dredge up of lower entropy gas, but there is no evidence
of any AGN-inflated cavities external to the filaments. In fact, the
filaments are confined to the regions between the largest X-ray cavities,
which were likely inflated by an AGN outburst about 13~Myr ago.  Another
possible method for dredging up lower entropy gas is through AGN-induced
turbulence.  Even though there are no obvious X-ray cavities exterior to
any of the X-ray or H$\alpha$ filaments in NGC 5044, the buoyantly rising cavities
may drive turbulence throughout the central region.  In NGC 5044, it is
possible that the X-ray filaments are uplifted lower entropy gas within the
turbulent eddies.

The kinetic energy within any AGN-induced turbulence will eventually
be dissipated as heat at a rate per unit volume given by

\begin{equation}
\Gamma_{turb} = c_2 \rho_g { {u^3} \over {l} }.
\end{equation}

\noindent
Here $\rho_g$ is the gas density, $u$ is the turbulent velocity,
$l$ is the length scale of the eddies and $c_2=0.42$ (Dennis \&
Chandran 2005). To balance radiative cooling within the central
2~kpc (the region covered by ALMA) by the dissipation of turbulent
kinetic energy requires a turbulent velocity of

\begin{equation}
u = 150~\left( { {l} \over {1~\rm{kpc}} } \right)^{1/3}~\rm{km}~\rm{s}^{-1}.
\end{equation}

\begin{deluxetable}{cccccc}
\tablecaption{Properties of the X-Ray Filaments}
\tablehead{
 \colhead{Region} &
 \colhead{n$_e$} &
 \colhead{$\rm{M_{gas}}$} &
 \colhead{$\rm{t_{c}}$} &
 \colhead{$\rm{\dot M}$} &
 \colhead{S} 
 \\
 \colhead{} &
 \colhead{(cm$^{-3}$)} &
 \colhead{($\rm{M_{\odot}}$)} &
 \colhead{(yr)} &
 \colhead{($\rm{M_{\odot}~yr^{-1}}$)} &
 \colhead{(keV cm$^{-2}$)} 
}
\startdata
X1     & 0.031       & $7.1 \times 10^6$  & $1.9 \times 10^8$   & 0.038                        & 7.0 \\
X2     & 0.024       & $8.1 \times 10^6$  & $2.8 \times 10^8$   & 0.029                        & 9.6 \\
X4     & 0.041       & $1.7 \times 10^7$  & $1.3 \times 10^8$   & 0.13                        & 5.3  \\
X5     & 0.040       & $1.7 \times 10^7$  & $1.4 \times 10^8$   & 0.12                        & 5.8  \\
X6     & 0.033       & $2.1 \times 10^7$  & $2.1 \times 10^8$   & 0.10                        & 8.3  \\
\enddata
\end{deluxetable}

\noindent
The only measurement of the turbulent velocity in
a cluster is the {\it Hitomi} XRS observation of the Perseus cluster, which
detected small-scale ($<$ 10~kpc) motions with a velocity
dispersion of 164~km~s$^{-1}$ (Hitomi Collaboration 2016).  This observed
level of turbulence is very similar to the level required to balance radiative
cooling in the core of NGC 5044.  It is also interesting to note that the
turbulent velocity required to balance radiative cooling is similar to
the observed velocity dispersion of the GMAs in NGC 5044. As noted
above, the observed velocities of the GMAs are inconsistent with
unimpeded gravitational infall over distances of more than 1~kpc. Thus, it is
possible that thermally unstable gas remains pinned to the ambient hot gas during
most of the cooling process, possibly by magnetic fields, as suggested
for the H$\alpha$ filaments in the Perseus cluster by Fabian et al. (2008),
and the observed velocity dispersion of the GMAs primarily reflects the
level of turbulence in the hot gas. This scenario does not require that the
GMAs are still pinned to the hot gas, only that the coldest gas recently
decoupled from the hot gas.

A summary of the basic properties of the X-ray filaments is given in
Table 5, excluding the filament X3, which has a complex
geometry.  The gas density and mass within each filament were computed
from the emission measure of the best-fit single-temperature model along
with an assumed geometry for each region: (1) cylindrical symmetry for
filaments X1, X2, X4, and X6; and (2) prolate spheroidal symmetry for filament X5.
Figure 13 shows that these filaments contain the majority of the
multiphase gas based on the Fe-L linewidth diagnostic.  The individual
masses of the X-ray filaments vary by a factor of three with an average mass
of $1.4 \times 10^7~\rm{M_{\odot}}$ and a total mass of
$7.0 \times 10^7~\rm{M_{\odot}}$.  The cooling times are also very similar
with an average cooling time of $2 \times 10^8$~yr.
The mass-cooling rates within the filaments vary
between 0.04-0.13$\rm{M_{\odot}}$~yr$^{-1}$ with a total mass-cooling rate
of 0.54~$\rm{M_{\odot}}$~yr$^{-1}$.  The inclusion of gas cooling
within the X3 filament would only increase this estimate by about 20\%.
It is interesting to note that the combined mass-cooling rate within the
multiphase gas detected with the Fe-L linewidth diagnostic is
approximately 10\% of the classical mass-cooling rate within 8~kpc (within
which the cooling time is shorter than 1~Gyr).

For comparison, the mean mass of the 24 GMAs detected
by ALMA is $\sim 10^6~\rm{M_{\odot}}$. Thus, each filament is capable of
cooling and fragmenting into tens of molecular structures with masses comparable
to those observed.  The combined mass of the 24 molecular
structures detected in the ALMA data is $5.1 \times 10^7 \rm{M_{\odot}}$, but
the integrated CO(2-1) flux in the IRAM 30m observation of NGC 5044
is approximately three times the integrated CO(2-1) flux of the 24 molecular
structures detected in the ALMA data.  The discrepency between the single-dish
and interferometric observations is probably due to the presence of
diffuse emission on scales greater than the scales sampled by the
ALMA observation.  Based on the CO(2-1) IRAM 30m observation of
NGC 5044, the total molecular mass in NGC 5044 is approximately
$1.5 \times 10^8 \rm{M_{\odot}}$.
Assuming the molecular gas is supplied by gas cooling
within the multiphase filaments, the supply time for the molecular
gas is $\sim 3 \times 10^8$~years.

The star formation rate in NGC 5044 is 0.07~$\rm{M_{\odot}}$~yr$^{-1}$
(Werner et al. 2014), which is approximately 13\% of the mass-cooling rate,
or equivalently, $\rm{ \dot M_c} / \rm{ \dot M_{SFR}} \approx 7$.
For comparison, Rafferty et al. (2006) derived mass-cooling rates for a sample
of CDGs in cooling flows through a spectroscopic analysis of ACIS data
using the isobaric cooling flow model MKCFLOW and found an average
value of $\rm{ \dot M_c} / \rm{ \dot M_{SFR}}$ of about four, but with values up to 20 for
systems with low rates of star formation like NGC 5044. Molendi et al. (2016) recently
presented a spectroscopic analysis using the MKCFLOW model for a sample
of BCGs with very high star formation rates, and in all cases, only
obtained upper limits on the mass-cooling rates and $\rm{ \dot M_c} / \rm{ \dot M_{SFR}}$
values as low as 0.1.  Molendi et al. discuss several possibilities for their
result, including the possibility of a delay between the peak mass
cooling and star formation rates.

Within molecular clouds, star formation is suppressed by magnetic
fields, turbulence, and stellar feedback (i.e., supernovae and stellar winds),
which prevents most of the molecular
gas from forming stars.  Assuming that one Type II supernova (SNe~II)
is produced for every 100~$\rm{M_{\odot}}$ consumed into stars and that
every supernova produces $10^{51}$~erg, we derive a SNe~II heating rate of
$2 \times 10^{38}$~erg~s$^{-1}$ in NGC 5044. This SNe~II heating
rate can only reheat the observed molecular gas up to $\sim 10^4$~K. Thus, SNe~II
heating, by itself, is not sufficient to reheat the molecular gas that is not
consumed into stars up to X-ray emitting temperatures.  There must therefore
be a two-stage feedback mechanism operating within the center of NGC 5044
where molecular gas that is not consumed into stars is reheated to
$\sim 10^4$~K by SNe~II and then some of this gas mixes with the hot gas
and is reheated to $\sim 1$~keV by AGN-feedback, while some gas re-cools
and forms a subsequent generation of molecular structures.

Our deep {\it Chandra} observation of NGC 5044 shows that the
central AGN has undergone only two outbursts over the past $10^8$ years
based on the presence of two pairs of nearly bipolar X-ray cavities.
In David et al. (2009) we argued that the AGN had undergone multiple
weak outbursts in the recent past.  This conclusion was based on the view that
all of the central depressions in the X-ray surface brightness were
AGN-inflated cavities.  A more detailed analysis of the deeper
{\it Chandra} data shows that these apparent depressions are simply
regions of average surface brightness located between X-ray
bright filaments.

\section{Summary}

Our main focus in this paper has been to identify the hot
gas responsible for the production of the observed
atomic and molecular gas in the center of NGC 5044.  Based on the Fe-L linewidth
technique, most of the multiphase thermally unstable gas in NGC 5044
is contained within the X-ray filaments.  Unlike some other groups and
clusters, which show obvious connections between cooler gas and
AGN-inflated cavities, the X-ray filaments in NGC 5044 are predominately
located between the largest X-ray cavities.
Given the existing data, it is difficult to distinguish between
in situ cooling or enhanced cooling within low-entropy gas that
has been dredged up by AGN-induced turbulence without more detailed
information about the kinematics and abundances of the gas within
the X-ray filaments.

While the cooling times, entropy, and free-fall times of the gas within the
X-ray filaments are very similar, the deep {\it Chandra} observation
of NGC 5044 shows that the X-ray spectra and association with H$\alpha$ emission
vary significantly between the filaments.  The variety of observed properties
among the X-ray filaments can be explained as an evolutionary sequence.
Of the six X-ray filaments, three do not show evidence of a soft X-ray
excess or any correlation with H$\alpha$ filaments.  It is likely
that these three filaments just recently began to cool out of the hotter ambient gas.
One X-ray filament exhibits a soft X-ray excess, but only contains some
diffuse H$\alpha$ emission, while another X-ray filament shows a strong
soft X-ray excess and is coincident with the brightest H$\alpha$ filament.
Thus, there is a general evolutionary sequence from the production of a cool
X-ray filament to the development of a soft X-ray excess to the production of
atomic gas.  However, even in the most evolved filament, X4, there is an
offset between the coolest X-ray emitting gas and the brightest portions of
the H$\alpha$ filament.

Our cycle 0 ALMA observation only covered the central 2~kpc of NGC 5044, so we
cannot make a detailed comparison between the X-ray filaments and
the molecular gas, but the observed low velocities of the molecular structures
suggest that most of the cooling took place while the gas
was prevented from freely falling through the gravitational
potential of NGC 5044, possibly as a result of the effects of drag or magnetic support.
The offset between the X4 and H$\alpha$ filaments shows that
the cooling gas eventually detaches from the hotter gas.  Without an extended 
free-fall period, the observed velocity dispersion of the molecular structures,
120~km~s$^{-1}$, should reflect the level of turbulence in the hot gas
from which it originated.  The turbulent velocity required to balance radiative
cooling with the dissipation of turbulent kinetic energy within the central 2~kpc of NGC 5044
is 150~km~s$^{-1}$. This calculation certainly suggests that the
dissipation of AGN-induced turbulence can prevent the bulk of hot gas from
cooling within the central region of NGC 5044.  As noted above, the H$\alpha$
and [CII] linewidths are significantly greater than the CO linewidth, but the
linewidths for the warmer gas are computed over a larger region.  We were awarded
a cycle 4 ALMA observation that will mosaic the central 5 by 5~kpc region of NGC 5044,
from which we can make a more direct comparison between the hot, atomic,
and molecular gas.

The mass-cooling rate within the X-ray filaments is approximately
10\% of the classical mass-cooling rate and eight times the observed
star formation rate.  Based on the mass-cooling rate within the X-ray filaments,
the molecular gas can be supplied in approximately $3 \times 10^8$ years.
The discrepancies between the classical mass-cooling
rate and the star formation rate can only be reconciled with a
two-stage feedback mechanism in which AGN-heating prevents
approximately 90\% of the hot gas from cooling to molecular gas, and stellar
feedback prevents 80-90\% of the molecular gas from forming stars.  The required
star formation efficiency in NGC 5044 is typical of that observed in star forming galaxies.
An estimate of the Type II supernova heating rate based on the observed
star formation rate shows that stellar feedback can only heat the molecular
gas not consumed into stars to about $10^4$~K.  Further heating by
AGN-feedback is required to reheat any residual molecular gas to X-ray
emitting temperatures.

This work was supported in part by NASA grant GO5-16135X.
We would like to acknowledge some very informative
conversations we had with Paul Nulsen, Grant Tremblay,
and Sebastian Heinz concerning the AGN-feedback mechanism
and the production of cold gas in cooling flows.

\end{document}